# From Subgroups to Population Composition: A Transportability Approach to Effect Heterogeneity

**Authors:** Michael Cheung[1,2], Candus Shi[1], Kara E Rudolph[3], Valérie Garès[4], Caroline A Thompson[5], Tarik Benmarhnia[1,6]

**Author Affiliations**

[1]Scripps Institution of Oceanography, University of California San Diego, La Jolla, CA, USA

[2]Francis I. Proctor Foundation and the Department of Ophthalmology, University of California, San Francisco, San Francisco, CA, USA

[3]Department of Epidemiology, Mailman School of Public Health, Columbia University, New York, NY, USA

[4]Univ Rennes, INRIA, CNRS, IRMAR-UMR 6625, IRSET, Rennes, France

[5]Department of Epidemiology, Gillings School of Global Public Health, The University of North Carolina at Chapel Hill, Chapel Hill, NC, USA; Lineberger Comprehensive Cancer Center, University of North Carolina at Chapel Hill, Chapel Hill, NC, USA

[6]Irset Institut de Recherche en Santé, Environnement et Travail, UMR-S 1085, Inserm, University of Rennes, EHESP, Rennes, France

*Corresponding author. Email: tbenmarhnia@ucsd.edu

**Abstract**
Identifying heterogeneous populations across which exposure effects vary is essential for transportability applications, cost-benefit analyses, and intervention prioritization. Traditional methods for heterogeneity analyses rely on parametric regression with prespecified subgroups, which may fail to capture complex patterns of effect modification. While recent data-adaptive methods improve high-dimensional heterogeneous effect prediction, they add methodological complexity to analyses and may offer limited insight into key drivers of heterogeneity. In this paper, we propose a novel, conceptual approach for heterogeneity analyses that considers how exposure effects would differ in populations with different compositions by modeling the population-level effect surface as a function of the distribution of effect modifiers. The approach consists of three steps: i) selecting confounders and effect modifiers based on prior knowledge (or alternatively using data-adaptive methods to learn effect modifiers), ii) estimating exposure effects in hypothetical populations with different effect modifier prevalences using transportability methods, and iii) modeling the estimated effects as a function of prevalence values. This approach provides two types of outputs: estimation of the change in the population-level exposure effects attributable to increases in effect modifier prevalence and ranking of effect estimates across multiple effect modifiers and prevalences to identify population characteristics most strongly associated with differential vulnerability. We demonstrate the approach using Demographic and Health Surveys data to examine heterogeneous effects of drought on child stunting and provide a Shiny application to implement this approach in any setting.

## 1 Introduction

A primary interest of epidemiologists and other quantitative social researchers involves understanding how and why the effect of an exposure varies across populations – not only identifying subgroups with heterogeneous exposure effects, i.e. the exposure effect varies in magnitude or direction across subgroups, but also characterizing the relationship between population characteristics and effect magnitude. Discrepancies in effects across populations arise notably due to differences in the distributions of effect modifiers or mediators between populations (Rudolph and Díaz, 2022a), which alter the average effect of an exposure on the outcome between those populations. Identification of populations with heterogeneous exposure effects is important for generalizability, accurate cost-benefit analysis, and subsequent intervention allocation and prioritization. Moreover, it is particularly crucial to identify vulnerable populations who are harmed by an exposure but may be obscured by an on-average null or protective effect (Rudolph and Díaz, 2022b).

Traditional methods for heterogeneity analyses involve parametric regression modeling with prespecified subgroups based on prior knowledge. These typically include two main approaches: i) incorporating an interaction term in a multivariable model; ii) conducting stratified analyses coupled with a heterogeneity test. In the first approach, heterogeneity is assessed by the interaction between the exposure and the variable(s) that constitute the relevant subgroups. In the second approach, subgroup effects are estimated by separate models and compared using a hypothesis test such as Cochran's Q test (Kaufman and MacLehose, 2013). While simple to implement, these methods are not optimal when multiple effect modifiers are of interest (VanderWeele et al., 2019) – manual specification of effect modifiers, relevant covariates, and the model form often ignores the complex, high-dimensional heterogeneous relationships that are more realistic in real-world data.

In the last decade, some methods have been proposed to improve heterogeneous exposure effect prediction accuracy through a variety of nonparametric techniques. *Meta-learner*s (Kennedy, 2023; Künzel et al., 2019; Nie and Wager, 2021), which comprise classes of algorithms that deconstruct heterogeneous effect estimation into steps that can be accomplished with separate predictive models, have been shown to perform well in a variety of settings (Chernozhukov et al., 2024; Dorie et al., 2019). These approaches are more capable of modeling high-dimensional heterogeneous effects than traditional methods but add methodological complexity that makes them more challenging to implement and communicate to external audiences. Moreover, when the purpose of an analysis is identifying and understanding key drivers of heterogeneity (to define actionable vulnerable populations), high-dimensional effect estimates may offer limited insight. Variable importance metrics and methods akin to feature selection such as LASSO and Classification and Regression Trees (CART) have been suggested to highlight covariates that play large roles in heterogeneous effect estimation (Cheung et al., 2025; Hines et al., 2025). More recently, methods have been designed to directly identify heterogeneous subgroups (Bargagli-Stoffi et al., 2024; Nianogo et al., 2025; Wang et al., 2024; Zhao et al., 2022). These are promising tools for heterogeneity analyses but have seen limited use in applied settings.

Overall, the available methods for heterogeneity analyses primarily focus on estimation of heterogeneous exposure effects and identification of heterogeneous subgroups as a means of characterizing effect modification. In this paper, we put forward a conceptual approach for heterogeneity analyses that address how exposure effects would differ in populations with different compositions, without requiring pre-specified subgroups or parametric assumptions. We build on recently proposed approaches for heterogeneity analyses and consider population-level exposure effects and effect modifier prevalences, that is, how the overall effect in a population changes as the proportion of individuals with a given characteristic increases or decreases.

Specifically, this approach provides two informative results: 1) estimation of the change in the population-level exposure effect attributable to an increase in the prevalence of an effect modifier (considering any functional form) and 2) a ranking of effect estimates across multiple effect modifiers and prevalences to identify the population characteristics most strongly associated with differential vulnerability. This approach does not focus on optimizing or improving the accuracy or efficiency of exposure effect estimation or effect modifier identification. Rather, the primary aim is to provide a practical conceptual approach for heterogeneity analyses as an alternative to conventional methods. While meta-learners and data-adaptive approaches provide high-dimensional effect estimation, this approach provides low-complexity visualization of the population-level exposure effect modification that is interpretable and outlines actionable vulnerable subgroups. We also include a *Shiny* application (Chang et al., 2012) in R (R Core Team, 2023) as a user-friendly alternative for researchers to implement the approach in their own work (https://github.com/benmarhnia-lab/heterogeneity_resampling_approach).

In section 2, we first briefly discuss the notation and assumptions used throughout the paper, followed by a detailed description of the approach. In section 3, we demonstrate its application in relation to the heterogeneous effects of droughts on child stunting using the Demographic and Health Surveys (DHS) data as a case study. Section 4 concludes with a discussion.

## 2 Methods

### *2.1 Method overview*

This section describes the proposed approach for heterogeneity analyses. The approach is comprised of a 3-step process detailed in the following subsections. In brief, the first step involves selecting the appropriate confounders and effect modifiers based on prior knowledge of the data and research question. In the second step, population-averaged exposure effects are estimated in hypothetical populations with different effect modifier distributions. The third step consists of modeling the estimated exposure effects as a function of the prevalence values to quantify the relationship between the effect modifier and the population-level exposure effect.

### *2.2 Terminology and assumptions*

We first establish the notation used to define terms referenced throughout the paper. Following the potential outcomes framework, let $Y(A)$ denote the potential outcome under the exposure $A \in \{0,1\}$ and $L$ be the covariate values. Let $(Y_i, A_i, L_i)$ be the observed data for unit $i = 1, \dots, n$. The effect of an exposure on a given outcome in a population is measured by the average treatment effect (ATE). Using the potential outcomes notation, the ATE is defined as $\mathbb{E}[Y(1) - Y(0)]$.

We assume the standard causal identification assumptions of consistency ($Y = AY(1) + (1 - A)Y(0)$), positivity ($P(A = a|L = l) > 0$ for all $l$ with $P(\mathrm{L} = l) \neq 0$), and conditional exchangeability ($Y(1), Y(0) \perp A|L$). Additionally, we require the analogous set of transport assumptions to estimate the exposure effect in hypothetical populations (Degtiar and Rose, 2023). First, let $E$ be the set of effect modifiers and $R \in \{0,1\}$ be the indicator for the study sample ($R_i = 1$ if unit $i$ is in the study sample, $R_i = 0$ if unit $i$ is in the hypothetical population). Consistency (and the stable unit treatment value assumption) requires that the treatment version is the same across the study sample and hypothetical populations, and there is no interference between units. Positivity of selection requires that within each stratum of effect modifiers, the probability of being in the study sample is non-zero, or alternatively, any stratum of the effect modifiers that exists in the hypothetical population must also have representation in the study sample ($P(E = e|R = 0) > 0 \Rightarrow P(E = e|R = 1) > 0$). Conditional exchangeability of the study sample and hypothetical population units is also required – the potential outcomes are independent of selection into the hypothetical population, conditioned on the observed effect modifiers and covariates ($Y(1), Y(0) \perp R|E, L$). Notably, this assumes no unmeasured effect modifiers, as well as other correlated covariates that differ in distribution between the study sample and hypothetical population. Last, to attribute changes in the population-level exposure effect across hypothetical populations solely to changes in the effect modifier distribution, one must assume there are no mediators on the causal pathway between the exposure and outcome that differ in distribution between the study sample and the hypothetical population. If an effect modifier is correlated with a mediator, fixing the effect modifier distribution at different prevalences will in turn shift the mediator's marginal distribution. In this case, the change in the estimated exposure effect across prevalence values will reflect both effect modification and a change in the natural indirect effect attributable to the mediator.

### *2.3 Step 1: Select confounders and effect modifiers*

The first step involves the researcher selecting an appropriate set of confounders and effect modifiers. It is crucial for the researcher to select confounders that close back door paths between the exposure and outcome, while carefully excluding variables that would otherwise induce collider stratification bias (Greenland, 2003). Higher-dimensional sets of effect modifiers may also be considered (i.e. joint effect modifier interactions) which our approach handles, but it is worth mentioning that interpretability is reduced with increasing dimensionality.

In the absence of prior knowledge of effect modifiers, meta-learners or other data-adaptive methods may be used to identify effect modifiers (Bargagli-Stoffi et al., 2024; Nianogo et al., 2025; Wang et al., 2024; Zhao et al., 2022). Additionally, descriptive summaries of the data may be presented in this step. Potential violations of positivity that may limit analyses will be reflected in these summaries.

### *2.4 Step 2: Estimate the exposure effect in hypothetical populations with different effect modifier prevalences*

The next step involves estimating the average exposure effect in hypothetical populations with different prevalences of an effect modifier. Implementing this procedure across a range of target prevalences, for example, the set of quintile values (0%, 20%, 40%, 60%, 80%, 100%), can provide an informative set of population-level effect estimates that together describe how the exposure effect varies as the effect modifier distribution changes. This step is agnostic to the specific estimation method, and can be accomplished using transportability weighting, outcome model standardization, or doubly robust methods such as targeted maximum likelihood estimation (TMLE).

*Transportability weighting:* Under this approach, the data are transported to hypothetical populations using weighting methods, which construct weights that rebalance the study sample to reflect the target effect modifier prevalence. Common examples include inverse odds weighting and entropy balancing (Westreich et al., 2017; Josey et al., 2021). Under inverse odds weighting, weights for unit $i$ are defined as:

$$w_i = \begin{cases} \frac{P(R_i = 0 | E_i = e_i)}{P(R_i = 1 | E_i = e_i)} \times \frac{P(R_i = 1)}{P(R_i = 0)}, & R_i = 1 \\ 0, & R_i = 0 \end{cases}$$

where the quantity $P(R_i = 1 | E_i = e_i)$ can be estimated by logistic regression. Entropy balancing consists of solving the minimization problem

$$\text{minimize} \sum_{\{i:R_i=1\}} w_i \log w_i - w_i$$

$$\text{subject to} \sum_{\{i:R_i=1\}} w_i E_i = p, \sum_{\{i:R_i=1\}} w_i = 1$$

where $p$ is the desired effect modifier prevalence in a hypothetical population. A confounding-adjusted exposure effect model is then fit using these weights.

*Outcome model standardization (g-computation):* An outcome model $\hat{E}[Y \mid A, E, L]$ is fit on the original study sample where $E$ is an effect modifier whose distribution we wish to set to the target prevalence $p$. To estimate the exposure effect in a hypothetical population with effect modifier prevalence $p$, the effect modifier is set to the target prevalence value for all individuals while all other covariate values $L_i$ are held at their observed values. Predictions $\hat{E}[Y \mid A = 1, E = p, L_i]$ and $\hat{E}[Y \mid A = 0, E = p, L_i]$ are then averaged over the full study sample to obtain the estimated ATE.

*Doubly robust estimators:* Doubly robust estimators, including augmented inverse probability weighted estimators and targeted maximum likelihood estimation (TMLE), combine outcome modeling with transportability weighting, providing protection against misspecification of either model (Josey et al., 2022; Lee et al., 2023; Rudolph and Van Der Laan, 2017). These estimators are consistent if either the outcome model or the weight model is correctly specified and are semi-parametrically efficient under correct specification of both.

Regardless of the chosen estimation method, this step may be repeated over resampled data (i.e., bootstrapping) to obtain confidence intervals for the exposure effect at each prevalence value. Positivity of selection should be assessed at each target prevalence, particularly at extreme values (0% and 100%), where weight distributions and covariate overlap should be inspected before interpreting results.

*2.5 Step 3: Model exposure effects as a function of prevalence values*

Once exposure effects are estimated for different effect modifier prevalences, they can be ranked by magnitude to descriptively identify heterogeneous and vulnerable populations with increased risks compared to the ATE. These estimates can be modeled in a second-stage regression as a function of the prevalence values to quantify the relationship between the population-level exposure effect and distribution of an effect modifier. The regression can fit any functional form that best characterizes the relationship, but it is important to balance interpretability of the model estimates and plausibility of the model assumptions. For example, an ordinary least squares model provides a simple interpretation of the change in the exposure effect due to an increase in the effect modifier prevalence but stipulates a strong assumption of linearity.

*2.6 Shiny application framework*

We provide a *Shiny* application (Chang et al., 2012) for researchers to implement this approach with their data. *Shiny* is a framework for developing interactive applications that can perform analyses without user knowledge of R or other statistical programming languages. The code is stored at https://github.com/benmarhnia-lab/heterogeneity_resampling_approach and the corresponding markdown files provide instructions for running the application.

## 3 Application: Demographic and Health Surveys Data

We demonstrate the application of the approach using the Demographic and Health Surveys (DHS) data. The DHS is a multi-country survey collecting information about health-related issues and status of women in reproductive age (ages 15 to 49) and their children under 5 years of age in low- and middle-income countries (LMIC). These include, but are not limited to, fertility, mortality, diseases, nutrition, and health-seeking behavior. More details regarding the data are provided in the appendix.

For our analysis, we selected drought as the exposure and stunted child growth as the outcome. There is strong evidence that droughts pose a serious threat to child health and nutrition in LMICs (Belesova et al., 2019; Cooper et al., 2019). However, the population groups most at risk have not been well studied. The extent to which climate shocks impact human health is likely to vary depending on the underlying level of population vulnerability.

The selected confounders are child sex, age, birth size, breastfed status, the mother's education level, single status, occupation, mass media consumption, rural residence, and wealth level. We considered three variables for effect modifiers: mother's education level, rural residence, and mass media consumption. These variables were chosen based on prior subject matter knowledge (Belesova et al., 2019) and for simplicity of demonstration. All variables are expressed as binary and missingness was removed such that all observations are complete. Table 1 shows the counts and prevalences of these variables in the data.

To adjust for confounding, we used stabilized inverse probability weighting, where the probability of exposure was estimated from a logistic regression model with all main effects of confounders. Exposure effects were weighted to the hypothetical populations of different effect modifier prevalences using inverse odds weighting and the set of quintiles was chosen as target prevalences. 95% confidence intervals for the exposure effects were estimated using 5000 bootstrap samples.

The estimate of the ATE (risk difference) in the original sample data is 0.014 (95% CI: [0.0096, 0.019]), indicating that out of every 1,000 children exposed to droughts, approximately 14 more stunting cases would occur compared to if those same children had not been exposed. Table 2 gives the estimated exposure effects for the hypothetical populations with different effect modifier prevalences ranked by magnitude. The largest exposure effect, a 3.3% increase in the risk of stunted child growth, belongs to the population with 100% prevalence of no maternal education. Moreover, the populations with at least 60% prevalence of no maternal education have larger exposure effects than the estimated ATE. Conversely, we see a diminished exposure effect for the populations with prevalence below 60% of rural residence and below 20% of mass media consumption.

**Figure 1** plots the estimated exposure effects across prevalence values for each effect modifier, and the relationship between the population-level exposure effect and effect modifier prevalence is visually distinct across the three modifiers. For mother's education and mass media consumption, the exposure effect increases with prevalence at a decelerating rate. For rural residence, the opposite pattern is observed, with the exposure effect increasing at an accelerating rate. To characterize these relationships, we fit a second-order polynomial regression of the estimated exposure effects on prevalence values for each effect modifier. The quadratic specification was chosen following visual inspection of Figure 1, which suggested nonlinear relationships for all three modifiers. However, researchers should assess whether a linear, quadratic, or alternative functional form best characterizes the relationship in their own data prior to model fitting. Table 3 gives the estimated regression formulas by effect modifier. For mass media consumption, the function describes an estimated average increase of 0.70% in the risk of stunted child growth when the prevalence increases from 0% to 20%. However, when the prevalence increases from 80% to 100%, the estimated average risk decreases by 0.13%. This reversal in the direction of the marginal change, captured by the quadratic fit, highlights the value of modeling the full effect surface rather than relying on a subgroup comparison from a conventional stratified analysis.

## 4 Discussion

Using the proposed approach, researchers can: 1) visualize effect modification due to an increase in the prevalence of an effect modifier in a population and 2) rank effect measures across multiple effect modifiers to identify which characteristics are most strongly associated with differential vulnerability across the population distribution. These functions go beyond the binary descriptions of conventional heterogeneity subgroup analyses and have implications for intervention assessment and prioritization in external populations. The approach is agnostic to the effect estimation method (and thereby agnostic to the effect scale and measure), which provides researchers the flexibility to use the most appropriate estimation model in their own analyses. By leveraging recent developments in the transportability literature, this approach does not require data outside of a study sample to investigate exposure effects in external populations. Rather than transporting to an observed external population, the same methods are applied to hypothetical populations defined entirely within the study sample by varying the prevalence of effect modifiers. This means that the full range of effect modification can be explored using only the data at hand, without requiring a separate target dataset. Different transport methods may also be used to estimate the exposure effect in the hypothetical populations, including weighting, outcome model standardization, and doubly robust approaches.

Using the DHS data, we found strong effect modification across the prevalences of mother's education level, mass media consumption, and rural residence. The estimated ATE in the original sample corresponds to a 1.4% increase in the risk of stunted child growth due to drought exposure. Populations with a prevalence of no maternal education above 60% had exposure effects exceeding this average, with those at 100% prevalence facing an estimated 3.3% increase in risk. In populations where the prevalence of no maternal education approaches zero (that is, nearly all mothers have at least some education), the estimated exposure effect becomes negative, suggesting a small protective association between drought and stunting in highly educated populations. This may reflect the capacity of educated mothers to buffer the nutritional consequences of drought through adaptive behaviors such as dietary substitution or access to health services. While this finding warrants cautious interpretation given the uncertainty in estimates at extreme prevalences due to potential positivity violations, it illustrates the value of modeling the full effect surface rather than relying on a single subgroup comparison, which might obscure such heterogeneity. Furthermore, findings such as these could inform intervention prioritization by identifying vulnerable populations with similar high risk prevalences of the effect modifying characteristic. For example, geographic targeting of nutrition programs to villages or districts where maternal education levels are low and drought risk is high could reduce the burden of stunting attributable to drought.

While the demonstration was restricted to three binary effect modifiers for simplicity, the approach generalizes to larger sets of effect modifiers and, with extensions to the current implementation, to continuous effect modifiers. In practice, researchers may consider a richer set of vulnerability indicators simultaneously, provided that positivity and interpretability are maintained. It is also important to note that drought is an area-level exposure, which typically involves a simpler confounding structure than individual-level exposures, and the DHS survey weights were not incorporated for ease of demonstration. Applied researchers should carefully specify confounding structures excluding colliders and incorporate appropriate adjustments such as survey weights during the exposure effect estimation step, as they would in conventional heterogeneity analyses.

While the framework we propose is novel in its general formulation for heterogeneity analyses, it is worth situating it relative to prior work that has employed related ideas in more constrained settings. Inoue et al. (2022) used a similar approach to investigate how living arrangement and race modify the relationship between intensive blood pressure control and cardiovascular events using data from clinical trial participants. Specifically, the authors considered the set of quintile prevalences for both effect modifiers (0%, 20%, 40%, 60%, 80%, and 100%) using inverse odds of weighting to generate weights for the hypothetical populations. This can be seen as a special case of our proposed approach applied to a randomized, experimental population. Moreno-Betancur et al. (2018) proposed a conceptual framework for defining estimands to measure the change in disease occurrence due to joint and individual distributional variations in effect modifiers. This framework is fundamentally different from the approach we propose but analogously provides structure to quantify the change in an exposure effect due to effect modifier distribution differences while adjusting for confounders.

This approach has several limitations. First, it does not improve upon the efficiency or prediction accuracy of heterogeneous exposure effect estimation. The approach draws on existing methods, transportability methods to estimate exposure effects in hypothetical populations, and optionally meta-learners or other data-adaptive methods to identify effect modifiers, and is therefore constrained by the performance and assumptions of those chosen methods. Second, the approach does not improve statistical power to detect heterogeneous effects, nor does it resolve empirical positivity violations, whether near or literal. These are known limitations of conventional heterogeneity analyses and are not unique to this approach. Poor precision due to limited data within effect modifier strata will propagate through the approach and be reflected in estimates of uncertainty. For instance, bootstrap confidence intervals will widen when resampled datasets frequently undersample sparse strata. Researchers should interpret effect estimates at extreme prevalence values with caution. Third, the assumption that no mediators on the causal pathway between the exposure and outcome differ in distribution between the study sample and the hypothetical population may not always hold. As discussed in Section 2.2, violations of this assumption mean that changes in the estimated exposure effect across hypothetical populations reflect not only effect modification but also changes in the natural indirect effect.

Heterogeneity analyses are vital to understanding the relationship between exposures and outcomes for different populations. This work puts forward a conceptual approach for heterogeneity analyses that provides a more thorough description of effect modification than conventional methods, specifically moving from binary subgroup comparisons to a continuous characterization of how the population-level exposure effect varies as a function of effect modifier prevalence. Several directions remain for future work. For example, simulation studies evaluating the finite-sample performance of the approach across a range of data-generating scenarios, varying sample sizes, degrees of effect modification, and levels of positivity violation, would help establish practical guidance for applied researchers. As climate shocks, social inequities, and other population-level stressors increasingly motivate targeted interventions, methods that characterize the full surface of effect modification rather than isolated subgroup comparisons will be essential for producing actionable epidemiological evidence.

**Tables and Figures**

Table 1: *Descriptive statistics*

| Variable | Overall, N = 345,499[1] | Stunted Child Growth | |
|---|---|---|---|
| | | **Not Stunted**, N = 212190 (61%)[1] | **Stunted**, N = 133309 (39%)[1] |
| *Exposure* | | | |
| Drought | 50,310 (15%) | 30,034 (14%) | 20,276 (15%) |
| *Effect Modifiers* | | | |
| Mother's Education - None | 174,528 (51%) | 99,828 (47%) | 74,700 (56%) |
| Mass Media Consumption - Yes | 156,872 (45%) | 105,636 (50%) | 51,236 (38%) |
| Residence - Rural | 249,258 (72%) | 142,971 (67%) | 106,287 (80%) |
| *Covariates* | | | |
| Child Breastfed - Never | 7,463 (2.2%) | 4,558 (2.1%) | 2,905 (2.2%) |
| Child Sex - Male | 174,324 (50%) | 103,099 (49%) | 71,225 (53%) |
| Child Age - Under 2 | 127,378 (37%) | 84,656 (40%) | 42,722 (32%) |
| Single Mother - Yes | 27,128 (7.9%) | 16,348 (7.7%) | 10,780 (8.1%) |
| Agricultural Occupation - Yes | 161,724 (47%) | 87,741 (41%) | 73,983 (55%) |
| Childbirth Size - Small | 62,218 (18%) | 34,449 (16%) | 27,769 (21%) |
| Wealth - Poor | 158,259 (46%) | 83,998 (40%) | 74,261 (56%) |

[1]n (%)

*All variables recorded as binary*

Table 2: *Estimated exposure effects ranked by estimate across effect modifiers and prevalences. Effects are expressed as risk differences.*

| **Effect Modifier** | **Prevalence** | **Estimate** | **Lower 95% CI** | **Upper 95% CI** |
|---|---|---|---|---|
| Mother's Education - None | 100 | 0.033 | 0.027 | 0.04 |
| Mother's Education - None | 80 | 0.027 | 0.022 | 0.033 |
| Residence - Rural | 100 | 0.022 | 0.016 | 0.027 |
| Mother's Education - None | 60 | 0.02 | 0.016 | 0.025 |
| Mass Media Consumption - Yes | 80 | 0.019 | 0.014 | 0.025 |
| Mass Media Consumption - Yes | 60 | 0.019 | 0.014 | 0.023 |
| Mass Media Consumption - Yes | 100 | 0.018 | 0.011 | 0.025 |
| Mass Media Consumption - Yes | 40 | 0.016 | 0.011 | 0.02 |
| Residence - Rural | 80 | 0.015 | 0.011 | 0.02 |
| *Average treatment effect in the study sample* | | 0.014 | 0.0096 | 0.019 |
| Mother's Education - None | 40 | 0.012 | 0.007 | 0.017 |
| Mass Media Consumption - Yes | 20 | 0.011 | 0.006 | 0.016 |
| Residence - Rural | 60 | 0.01 | 0.005 | 0.014 |
| Residence - Rural | 40 | 0.006 | 0 | 0.011 |
| Mass Media Consumption - Yes | 0 | 0.004 | -0.002 | 0.01 |
| Residence - Rural | 20 | 0.003 | -0.004 | 0.009 |
| Mother's Education - None | 20 | 0.002 | -0.003 | 0.008 |
| Residence - Rural | 0 | 0 | -0.007 | 0.008 |
| Mother's Education - None | 0 | -0.009 | -0.015 | -0.002 |

Table 3: *Second-stage polynomial regression formula*

*(*Exposure Effect $\sim \beta_0 + \beta_1 * \text{prevalence} + \beta_2 * \text{prevalence}^2$*)*

| **Effect Modifier** | **Regression Formula** |
|---|---|
| Mother's Education – None | $-0.009 + 0.001 * \text{prevalence} - 2 \times 10^{-6} * \text{prevalence}^2$ |
| Mass Media Consumption | $0.004 + 4 \times 10^{-4} * \text{prevalence} - 3 \times 10^{-6} * \text{prevalence}^2$ |
| Residence - Rural | $0.001 + 1 \times 10^{-4} * \text{prevalence} + 1 \times 10^{-6} * \text{prevalence}^2$ |

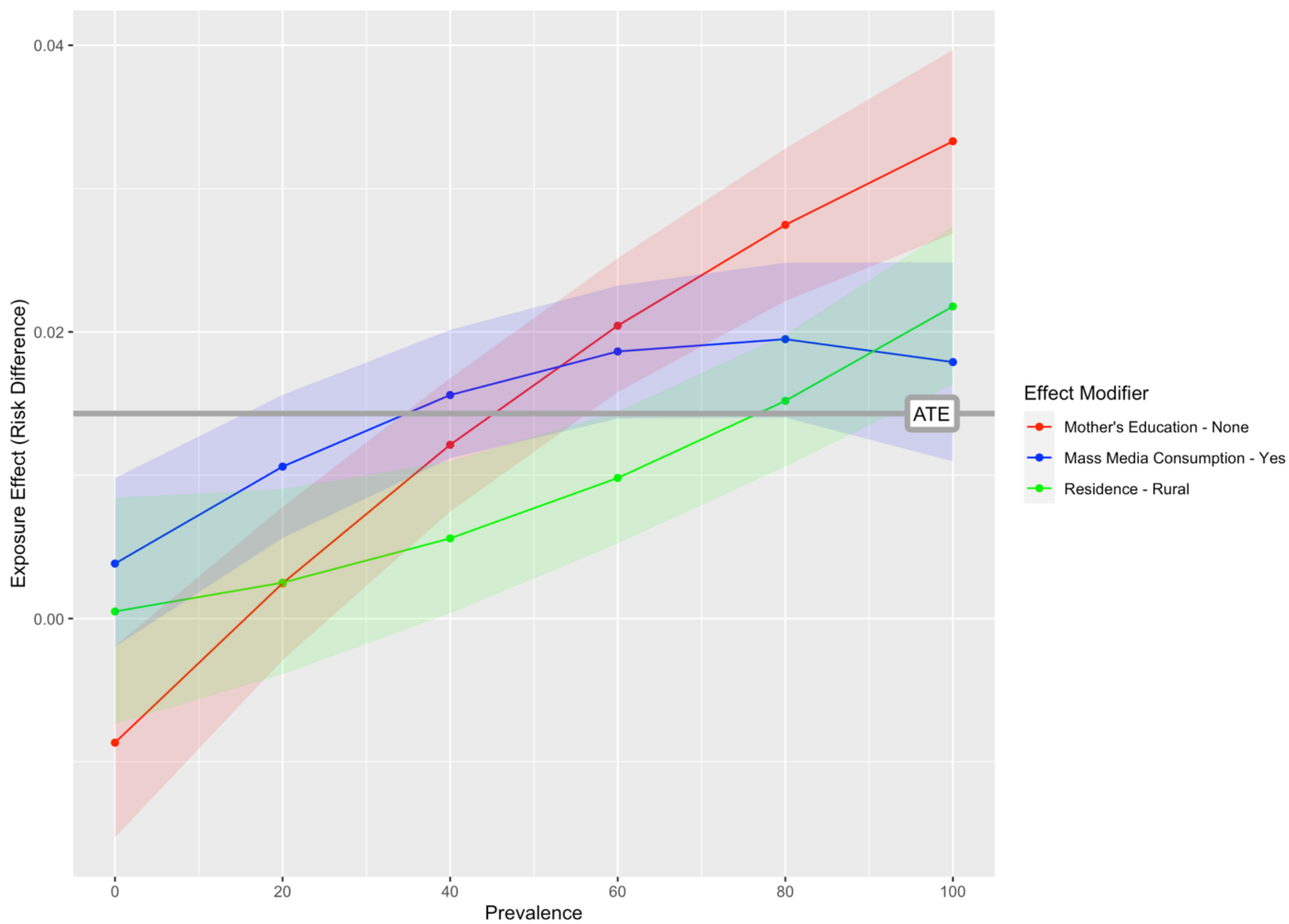

*Figure 1. Line plot of the estimated exposure effects in hypothetical populations across the prevalence quintiles of mother's education level, rural residence, and mass media consumption (estimates reported in table 2). 95% bootstrap confidence bands are shown for each effect modifier and the average treatment effect in the original study sample (0.014) is shown by the gray line.*

**Appendix**

The DHS surveys are repeated cross-sectional surveys that have been collected in over 90 low- and middle-income countries (LMIC) since the 1980s. These data focus on women in reproductive age (ages 15 to 49) and their children and cover a wide range of health-related issues, including fertility, mortality, diseases, nutrition, and health-seeking behavior. The surveys also include detailed socioeconomic information, such as household assets, urban or rural places of residence, and type of occupation, among other information. A two-stage cluster sampling process guarantees that the data collected is nationally representative. More recent survey rounds include global positioning system (GPS) information (latitude and longitude) for each primary sampling unit (PSU). A PSU is defined as a city block in an urban area and a village in a rural area. We were able to connect the survey data with high-resolution gridded climate data via the GPS information.

For our analysis, we restrict the data to surveys collected between 2000 and 2022 in sub-Saharan African countries. We focus on children under 5 years of age for whom detailed information was collected, including anthropometric measurements (weight and height), birth weight, and feeding practices. To reduce the risk of misreporting bias, we restrict the sample to children who live with their mothers and are usual residents at the place of interview. Our final sample consists of 345,499 children from 32 countries and 86 individual surveys.

Following standard practice, we construct a binary indicator for child stunting based on the children's height-for-age z-scores (HAZ). Children are classified as stunted if their HAZ score is more than 2 standard deviations below the World Health Organization's (WHO) growth standard median for children of the same age-group (World Health Organization, 2019). We dropped observations with implausible HAZ scores (below -6 and above 6). Low HAZ values reflect chronic undernutrition in children (stunted growth) (De Onis, 2017). It is a serious health concern since stunted growth has been associated with a range of negative outcomes such as diminished cognitive function, learning challenges and an increased risk of chronic health conditions (Black et al., 2013; Arthur et al., 2015; Victora et al., 2008; Poveda et al., 2021; Prendergast and Humphrey, 2014). Nearly a third of children in sub-Saharan African are stunted – some of the highest levels observed worldwide (*Levels and Trends in Child Malnutrition Child Malnutrition*, 2023).

In addition to children's anthropometric measurements, we retrieve information about the demographic and socioeconomic characteristics of the children and their households, including the mother's level of education (grouped into none or some education), the type of occupation of the household head (agricultural or non-agricultural), the location of the household (urban or rural), mother's consumption of mass media (including television, radio and newspaper), and household's relative wealth group (constructed based on standard DHS procedures) (Rutstein, 2015).

To assess the impact of climate shocks on the health status of LMIC children under 5, we merged the survey data with Standardized Precipitation Evapotranspiration Index (SPEI), a multi-scalar drought index that is commonly used to measure the intensity and spatial distribution of droughts. The SPEI index is computed using input monthly precipitation and potential evapotranspiration data from the Climatic Research Unit at the University of East Anglia (CRU

TS4) (Harris et al., 2020). We focus on identifying droughts during key agricultural periods. In particular, the main crop growing period for each PSU location is identified using high-resolution gridded data on the geographical distribution of crop areas (Monfreda et al., 2008) and crop calendar information (Sacks et al., 2010). Following standard practice, droughts are defined as crop-growing period SPEI values below -1. We focus on exposure to agricultural droughts during the infancy period (the first 12 months of life) since this is a period when most growth faltering is shown to occur (Alderman and Headey, 2018).

The degree to which agricultural droughts disrupt livelihoods and have an impact on human health depends on both individual and community-level factors, including dependence on agricultural produce, the capacity to diversify income, and access to information, to mention a few. Identifying population groups that are particularly vulnerable to climatic shocks is important for designing targeted interventions. Yet, few studies have looked into such disparities in drought-induced undernutrition, possibly due to the limitations of traditional methods of identifying effect modifiers.